\documentclass{article}
\usepackage{spconf,amsmath,graphicx}
\usepackage{amsmath} 
\usepackage{amssymb}  
\usepackage{amsthm}
\usepackage{amsfonts}
\usepackage{dsfont}
\usepackage{bm}
\usepackage{hyperref}
\usepackage{url}
\usepackage{multirow}
\usepackage{subcaption}

\newcommand{\bb}{\mathbf{b}}
\newcommand{\wb}{\mathbf{w}}

\newcommand{\Wb}{\mathbf{W}}

\newcommand{\Hb}{\mathbf{H}}

\newcommand{\Nc}{\mathcal{N}}

\newcommand{\RR}{\mathbb{R}}

\title{Image denoising with graph-convolutional neural networks}
\name{Diego Valsesia, Giulia Fracastoro, Enrico Magli\thanks{This research has been partially funded by the Smart-Data@PoliTO center for Big Data and Machine Learning technologies. We thank Nvidia for donating a Quadro P6000 GPU.}}
\address{Department of Electronics and Telecommunications -- Politecnico di Torino, Italy}
\begin{document}
%
\maketitle
\begin{abstract}
Recovering an image from a noisy observation is a key problem in signal processing. Recently, it has been shown that data-driven approaches employing convolutional neural networks can outperform classical model-based techniques, because they can capture more powerful and discriminative features. However, since these methods are based on convolutional operations, they are only capable of exploiting local similarities without taking into account non-local self-similarities. In this paper we propose a convolutional neural network that employs graph-convolutional layers in order to exploit both local and non-local similarities. The graph-convolutional layers dynamically construct neighborhoods in the feature space to detect latent correlations in the feature maps produced by the hidden layers. The experimental results show that the proposed architecture outperforms classical convolutional neural networks for the denoising task.
\end{abstract}
\begin{keywords}
Image denoising, Convolutional Neural Networks, Graph convolution
\end{keywords}

\vspace{-0.15cm}
\section{Introduction}
\label{sec:intro}
\vspace{-0.15cm}

Image denoising is a staple of the research on inverse problems, aiming to restore a clean image from a noisy observation, usually corrupted by additive white Gaussian noise. The key to the solution of this problem is to exploit prior knowledge about the structure of a natural image. The extensive literature on the topic has traditionally focused on developing increasingly refined hand-crafted models for natural images. Popular model-based algorithms include sparse representations \cite{elad2006image}, total variation methods \cite{rudin1992nonlinear,pang2017graph}, and methods based on non-local self-similarities such as BM3D \cite{dabov2007image}, or WNNM \cite{gu2014weighted}, which are among the most successful ones. However, recent work \cite{zhang2017beyond,lefkimmiatis2018universal,mao2016image,lehtinen2018noise2noise} has shown that a data-driven approach, whereby a convolutional neural network (CNN) is trained for the denoising task, can capture highly complex image priors without the need to handcraft them outperforming the best model-based algorithms. CNN-based approaches stack multiple convolutional layers interleaved by nonlinearities to create hierarchies of feature extractors. However, the main limitation is the local nature of the convolution operation, which is unable to increase the receptive field of a neuron-pixel to model non-local image features. This means that CNNs are unable to exploit the self-similar patterns that were proven to be highly successful in model-based methods. 

This paper presents a novel convolutional modeling by defining a new layer exploiting graph convolution, a generalization of the traditional convolution operation to deal with data that may not lie on regular grids, drawing from ideas in graph signal processing \cite{shuman2013emerging,Valsesia2019Sampling}. The proposed graph-convolutional layer allows to extract features that depend both on spatially-adjacent pixels, but also on spatially-distant pixels which nevertheless exhibit latent correlations by being close in the feature space. Notice that this approach exploits non-locality in a different manner with respect to algorithms based on global block matching (e.g., BM3D) because it builds a nonlinear hierarchy of filters with a non-local receptive field. While block-matching has a handcrafted notion of similarity (e.g., Euclidean distance between blocks in the noisy image), the proposed technique evaluates the receptive field from similarities in the latent space of the features constructed by the hidden layers of the network, thus greatly improving its expressive power.
As a final remark, our method can also be used in conjunction with a preprocessing stage based on wavelets \cite{bae2017beyond} or inside an iterative optimization scheme \cite{dong2018denoising}, which have been shown to improve denoising performance. In this paper, however, the core of a denoiser based on a graph-convolutional network is designed, while pre- or post-processing methods are left for future work.

\begin{figure*}[t]
    \centering
    \includegraphics[width=0.9\textwidth]{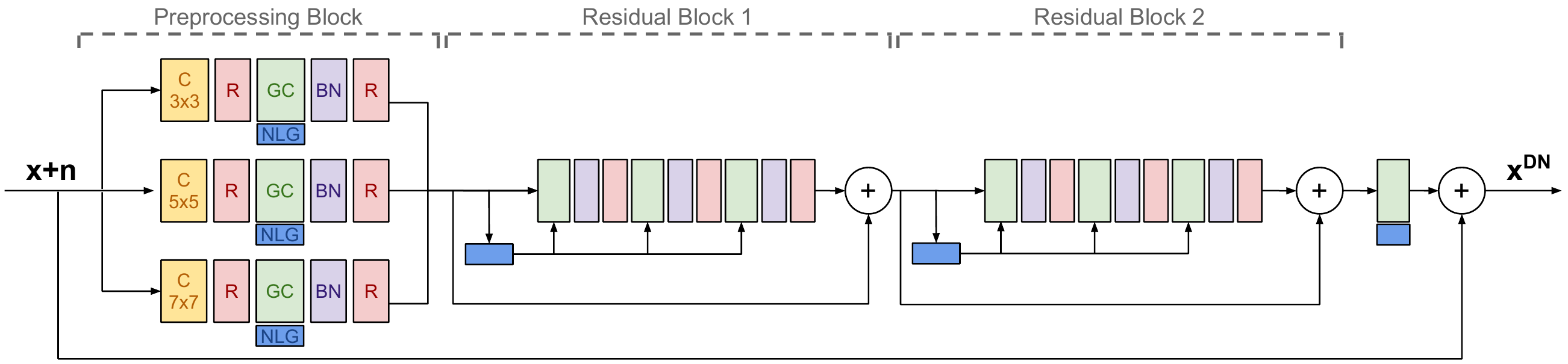}
    \vspace{-0.15cm}
    \caption{GraphCNN denoiser. Colored blocks are defined in the leftmost part. C: 2D convolution, R: leaky ReLU, GC: graph convolution (as in Fig.\ref{fig:gconv}), NLG: non-local graph construction (nearest neighbors in feature space), BN: batch normalization.}
    \label{fig:graphcnn}
    \vspace{-0.4cm}
\end{figure*}

\vspace{-0.3cm}
\section{Background}
\vspace{-0.2cm}
\label{sec:bkg}
\subsection{Graph convolution}
\vspace{-0.1cm}
In the proposed model, we employ the Edge Conditioned Convolution (ECC) presented in \cite{simonovsky2017dynamic}. This operation is defined using a spatial approach by performing weighted aggregations over a neighborhood.
Let us consider a layer $l$ with $N^l$ feature vectors of dimensionality $d^l$ and the corresponding graph $\mathcal{G}^l(\mathcal{V}^l,\mathcal{E}^l)$ where $\mathcal{V}^l$ is the set of vertices with $|\mathcal{V}^l|=N^l$ and $\mathcal{E}\subseteq \mathcal{V}^l\times\mathcal{V}^l$ is the set of edges. We assume that the edges of the graph are labeled, i.e. there exists a function $\mathcal{L}:\mathcal{E}\to \RR^s$ that assigns a label to each edge. In the following, we define the edge labeling function as the difference between the features of the two nodes of the edge, i.e. $\mathcal{L}(i,j)=\Hb_j^l-\Hb_i^l$. For each node $i$ of the graph $\mathcal{G}^l$, the convolution performs a weighted local aggregation of the feature vectors $\Hb_j^l \in \RR^{d^l}$ on the neighboring nodes $j \in \Nc_i^l$, where $\Nc_i^l$ is the neighborhood of node $i$. The weights of the local aggregation are defined by a fully-connected network $F^l:\RR^{d^l}\to\RR^{d^{l+1}\times d^{l}}$, which takes as input the 
edge labels and outputs the corresponding weight matrix $\bm{\Theta}^{l,ji} = F_{\wb^l}^l\left(\mathcal{L}(i,j)\right) \in\RR^{d^{l+1}\times d^{l}}$.   Hence, the convolution operation is defined as:
\vspace{-0.2cm}
\begin{align}
\label{eq:graph_conv}
%
\Hb_i^{l+1} \hspace*{-2pt} &= \hspace*{-2pt} \sigma\left(\sum_{j\in\Nc_i^l}\frac{F_{\wb^l}^l\left(\Hb_j^l-\Hb_i^l\right)\Hb_j^l}{|\Nc_i^l|}+\Wb^l\Hb_i^l+\bb^l\right)\hspace*{-1pt} \nonumber \\
&= \hspace*{-1pt}\sigma\left(\underbrace{\sum_{j\in\Nc_i^l}\frac{\bm{\Theta}^{l,ji}\Hb_j^l}{|\Nc_i^l|}}_\text{neighborhood}+\underbrace{\Wb^l\Hb_i^l}_\text{node}+\bb^l\right),
%
\end{align}
\vspace{-0.1cm}
where $\wb^l$ are the weights parameterizing network $F^l$, $\Wb^l\in\RR^{d^{l+1}\times d^{l}}$ is a linear transformation of the node itself, $\bb^l$ a bias, and $\sigma$ a non-linearity. It is important to note that the filter weights depend only on the edge labels. Therefore, two pairs of nodes with the same labels will have the same weigths. This behaviour is similar to weight sharing in classical CNNs.

\begin{figure}
    \centering
    \includegraphics[width=0.49\columnwidth]{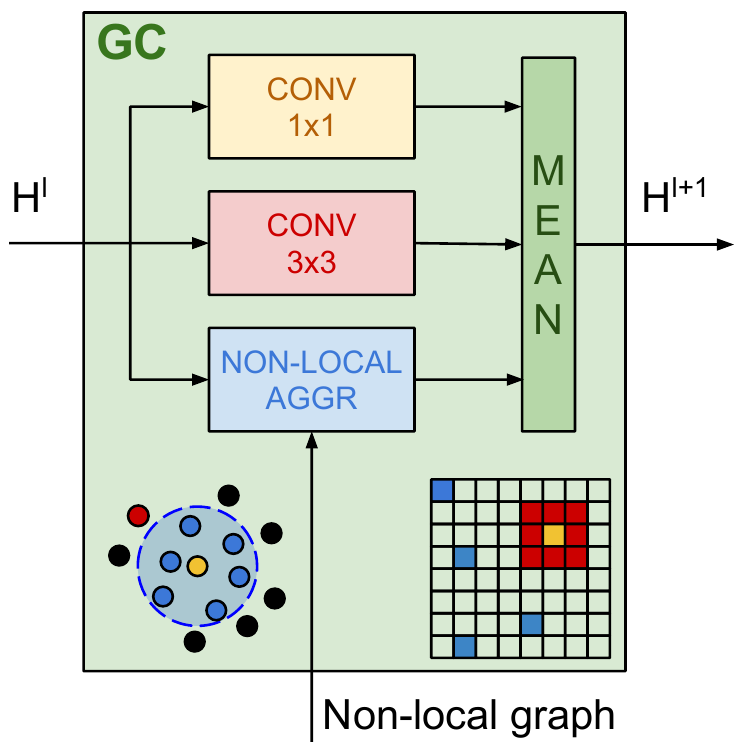}
    \vspace{-0.3cm}
    \caption{Graph-convolutional layer. Non-local aggregation block implements Eq. \eqref{eq:graph_conv}.}
    \label{fig:gconv}
    \vspace{-0.5cm}
\end{figure}

\vspace{-0.15cm}
\subsection{Image denoising with CNN}
\vspace{-0.15cm}
In the last years, several data-driven image denoising methods have been developed \cite{burger2012image,lefkimmiatis2018universal,lehtinen2018noise2noise,divakar2017image,zhang2017beyond}. Many of these methods employ convolutional neural networks (CNN) in order to learn more powerful and descriptive features \cite{zhang2017beyond,divakar2017image,lefkimmiatis2018universal}.
In this case, it has been shown that, rather than defining a CNN that outputs directly the denoised image, it is more effective to build a CNN that predicts the residual between the noisy observation and the clean image \cite{zhang2017beyond}. Using this approach, the CNN learns to progressively remove the clean image from the noisy observation. 
It is a common practice to train a CNN model for a fixed noise variance, even though works exist addressing the blind problem \cite{zhang2017beyond,chen2018image,lehtinen2018noise2noise}.

\vspace{-0.15cm}
\section{Proposed method}
\label{sec:method}
\vspace{-0.15cm}

In this section, we present the proposed architecture, called GraphCNN, to perform image denoising. An overview of the network is shown in Fig. \ref{fig:graphcnn}. At a high-level, the network is composed of a sequence of graph-convolutional layers followed by batch normalization and leaky ReLU nonlinearities. The first part of the network is composed of a preprocessing block with parallel branches, reminiscent of similar constructions in \cite{szegedy2015going} and \cite{divakar2017image}, whose goal is to extract features at multiple scales, by performing a classic convolution with three different filter sizes ($3\times 3$, $5\times 5$, $7\times 7$) followed by a graph convolution operation and finally concatenating the resulting features. The network also has several residual connections: notably, the input-output residual has been shown to be very effective for the denoising task \cite{zhang2017beyond} because it makes the network estimate noise features by progressively removing the clean image. The connection between the input and output of each residual block also improves gradient backpropagation.

\begin{table*}[]
\small
\setlength{\tabcolsep}{4pt}
\caption{Set12 PSNR (dB)}
\vspace{-0.2cm}
\begin{tabular}{lccccccccccccc}
\hline
       & Cman           & House          & Peppers        & Starfish       & Monarch        & Airplane       & Parrot         & Lena           & Barbara        & Boat           & Man            & Couple         & Avg             \\ \hline \hline
\multicolumn{14}{c}{$\sigma=15$} \\ \hline
BM3D \cite{dabov2007image}              & 31.91          & 34.93          & 32.69          & 31.14          & 31.85          & 31.07          & 31.37          & 34.26          & 33.10 & 32.13          & 31.92          & 32.10          & 32.372          \\ \hline
WNNM \cite{gu2014weighted}              & 32.17          & \textbf{35.13}          & 32.99          & 31.82          & 32.71          & 31.39          & 31.62          & 34.27          & \textbf{33.60} & 32.27          & 32.11          & 32.17          & 32.696          \\ \hline
OGLR \cite{pang2017graph}           & 31.36 & 34.88 & 32.31 & 30.70 & 31.26 & 30.46 & 30.87 & 33.97 & 32.54 & 31.58& 31.59 & 31.71 & 31.936 \\ \hline
DnCNN-S \cite{zhang2017beyond}           & \textbf{32.61} & 34.97          & \textbf{33.30} & 32.20          & 33.09          & 31.70          & 31.83          & \textbf{34.62} & 32.64          & \textbf{32.42} & \textbf{32.46} & \textbf{32.47} & 32.859          \\ \hline
\textbf{GraphCNN} & 32.58 & \textbf{35.13} & 33.27 & \textbf{32.42} & \textbf{33.25} & \textbf{31.84} & \textbf{31.89} & 34.57 & 32.84 & 32.41 & 32.42 & 32.40 & \textbf{32.917} \\ \hline 
\hline
\multicolumn{14}{c}{$\sigma=25$}             \\ \hline
BM3D \cite{dabov2007image}              & 29.45          & 32.85          & 30.16          & 28.56          & 29.25          & 28.42          & 28.93          & 32.07          & 30.71 & 29.90          & 29.61          & 29.71          & 29.969          \\ \hline
WNNM \cite{gu2014weighted}              & 29.64          & \textbf{33.22}          & 30.42          & 29.03          & 29.84          & 28.69          & 29.15          & 32.24          & \textbf{31.24} & 30.03          & 29.76          & 29.82          & 30.257          \\ \hline
OGLR \cite{pang2017graph}           & 29.11 & 32.65 & 30.02 & 28.29 & 29.16 & 28.10 & 28.76 & 31.95 & 30.35 & 29.59 & 29.47 & 29.49 & 29.744 \\ \hline
DnCNN-S \cite{zhang2017beyond}           & \textbf{30.18} & 33.06          & \textbf{30.87} & 29.41          & 30.28          & 29.13          & 29.43          & \textbf{32.44} & 30.00          & \textbf{30.21} & \textbf{30.10} & \textbf{30.12} & 30.436          \\ \hline
\textbf{GraphCNN} & 30.12          & \textbf{33.22} & \textbf{30.87} & \textbf{29.76} & \textbf{30.51} & \textbf{29.30} & \textbf{29.48} & 32.42          & 30.28          & 30.17          & \textbf{30.10} & 29.99          & \textbf{30.516} \\ \hline 
\hline
\multicolumn{14}{c}{$\sigma=50$}             \\ \hline
BM3D \cite{dabov2007image}              & 26.13          & 29.69          & 26.68          & 25.04          & 25.82          & 25.10          & 25.90          & 29.05          & 27.22 & 26.78          & 26.81          & 26.46          & 26.722          \\ \hline
WNNM \cite{gu2014weighted}              & 26.45          & \textbf{30.33}          & 26.95          & 25.44          & 26.32          & 25.42          & 26.14          & 29.25          & \textbf{27.79} & 26.97          & 26.94          & 26.64          & 27.052          \\ \hline
OGLR \cite{pang2017graph}           & 25.98 & 29.19 & 26.26 & 24.75 & 25.80 & 25.05 & 25.80 & 28.80 & 27.04 & 26.53 & 26.69 & 26.34 & 26.520 \\ \hline
DnCNN-S \cite{zhang2017beyond}           & \textbf{27.03} & 30.00          & 27.32 & 25.70          & 26.78          & 25.87          & \textbf{26.48}          & \textbf{29.39} & 26.22          & \textbf{27.20} & \textbf{27.24} & \textbf{26.90} & 27.178          \\ \hline
\textbf{GraphCNN} & 27.00 & 30.16 & \textbf{27.40} & \textbf{25.92} & \textbf{26.89} & \textbf{25.93} & 26.43 & 29.32 & 26.56 & 27.05 & 27.19 & 26.75 &  \textbf{27.217} \\ \hline 
\label{tab:results}
\end{tabular}
\vspace{-0.6cm}
\end{table*}

\vspace{-0.15cm}
\subsection{Graph-convolutional layer}
\vspace{-0.15cm}
The graph-convolutional layer is at the core of the proposed model. A schematic representation is shown in Fig. \ref{fig:gconv}. This layer extends the classical convolutional layer by aggregating the hidden-layer feature vectors of spatially-adjacent pixels as well as the hidden-layer feature vectors of spatially-distant pixels that are similar (nearest neighbors) in the feature space. The local features are aggregated using a classic $3\times3$ convolution while the non-local features are aggregated using the edge-conditioned graph convolution as defined in \eqref{eq:graph_conv}. The local and non-local contributions are then averaged to produce the output features.
The non-local pixels are chosen as the $k$-nearest-neighbor feature vectors in terms of Euclidean distance with respect to the feature vector of the current pixel within a search window of predefined size. Notice that the non-local selection is performed only at some hidden layers (as shown by the NLG block in Fig. \ref{fig:graphcnn}), with the two 3-layer residual blocks sharing the same non-local graph. This helps reducing complexity without compromising performance.

The role of the non-local graph in such residual architecture, whose goal is to successively remove the clean image from the noise features, is to identify the latent correlations in the feature space which are due to the residual image content rather than the uncorrelated noise.

We now analyse more in detail the role of the graph-convolution operation, implemented using the ECC of Eq. \eqref{eq:graph_conv}. Such definition provides two main contributions to the effectiveness of the algorithm but it is also a source of issues. The key to understanding it is the function $F$, which takes as input the difference between the feature vector of the current pixel and the feature vector of a non-local neighboring pixel and outputs the weight matrix used to transform the non-local feature vector before the aggregation. First, ECC can be called ``convolution'' because this function provides meaningful weight sharing under suitable stability assumptions: for a similar input difference, the output weight matrix should be similar. Second, differently from classical convolution this function enables a data-dependent aggregation because the weights depend directly on the relationships among feature vectors. 
The function $F$ was originally proposed \cite{simonovsky2017dynamic} to be implemented as fully-connected neural network with 2 layers. However, this leads to over-parameterization problems which negatively impact training. In order to understand this, let us focus on the last layer of the $F$-network: this layer is a dense layer with a $d^l$-dimensional input and a $d^{l+1}\times d^{l}$-dimensional output, meaning that the number of weights depends cubically on the number of features. This quickly leads to an excessively large number of parameters, causing vanishing gradients or overfitting. In order to reduce it, in this paper we propose to use a partially-structured matrix for the last layer of the $F$-network: instead of an unstructured matrix, we stack multiple row-subsampled circulant matrices. The effective number of parameters depends on i) how many subsampled matrices are stacked; ii) the row-subsampling factor of each of them. As an example, in our experiments we use $d^l=d^{l+1}=66$ and an unstructured matrix would require $66^3=287496$ parameters, while we stack $\frac{66^2}{3}$ partial circulant matrices with 3 rows each for a total of $66\frac{66^2}{3}=95,832$. Tradeoffs are possible by controlling how much structure one wants to enforce. Similar approaches to approximate fully connected layers have been studied in the literature \cite{wu2016compression, cheng2015exploration}.

\vspace{-0.25cm}
\section{Results}
\vspace{-0.2cm}

\begin{figure*}
    \centering
    \includegraphics[width=\textwidth]{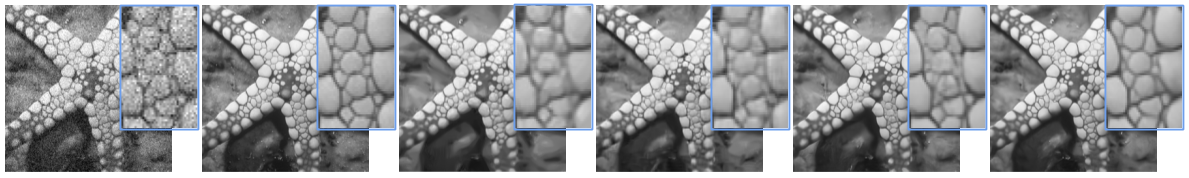}
    \vspace{-0.62cm}
    \caption{Denoising results for \textit{starfish}. Left to right: noisy ($\sigma=25$), original, OGLR (28.29 dB), BM3D (28.56 dB), DnCNN-S (29.41 dB), GraphCNN (\textbf{29.76 dB}).}
    \label{fig:04_denoised}
    \vspace{-0.2cm}
\end{figure*}
\begin{figure*}
    \centering
    \includegraphics[width=\textwidth]{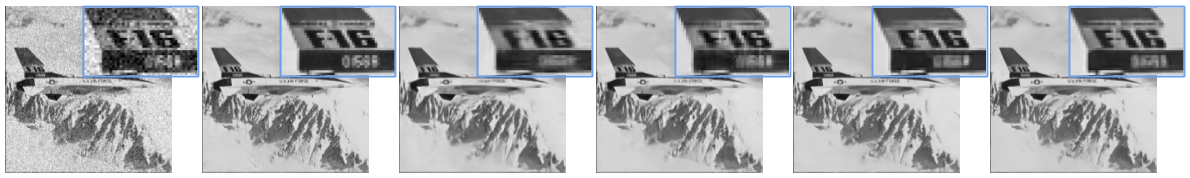}
    \vspace{-0.62cm}
    \caption{Denoising results for \textit{airplane}. Left to right: noisy ($\sigma=25$), original, OGLR (28.10 dB), BM3D (28.42 dB), DnCNN-S (29.13 dB), GraphCNN (\textbf{29.30 dB}).}
    \label{fig:06_denoised}
    \vspace{-0.4cm}
\end{figure*}

\begin{figure}
  \centering
        \includegraphics[width=0.24\columnwidth]{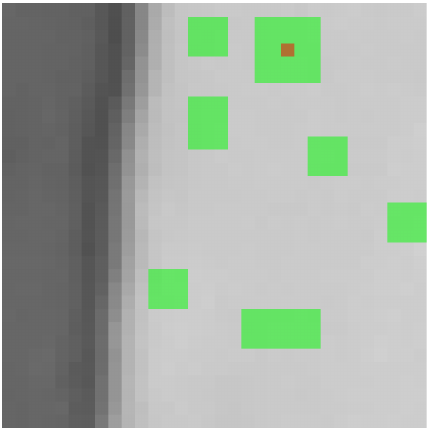}
        \includegraphics[width=0.24\columnwidth]{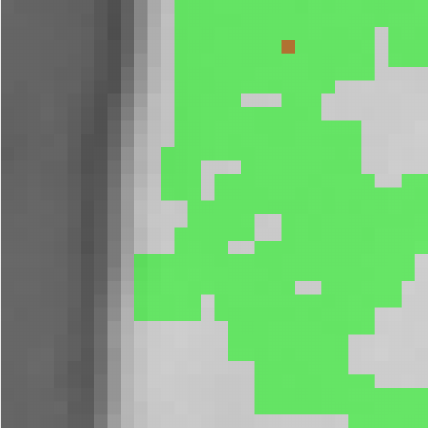}
        \includegraphics[width=0.24\columnwidth]{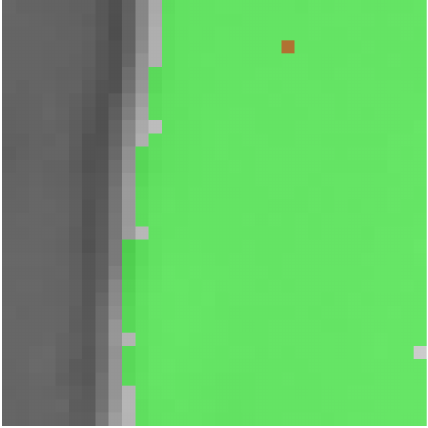}
        \includegraphics[width=0.24\columnwidth]{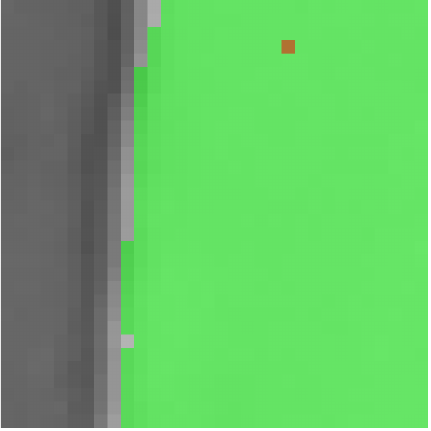}
 \vspace{-0.17cm}
 \caption{Receptive field (green) of a single pixel (red) for graph-convolutional layers 1-4 (ignoring the $5\times 5$ and $7\times 7$ multiscale branches for ease of representation).}
 \label{fig:recfield}
 \vspace{-0.4cm}
 \end{figure}

\label{sec:results}

In this section we perform an experimental evaluation of the proposed method, comparing the proposed method with several state-of-the-art denoising methods. We consider two model-based methods that exploit non-local priors (i.e., BM3D \cite{dabov2007image} and WNNM \cite{gu2014weighted}), a graph-based variational method (i.e., OGLR \cite{pang2017graph}) and a state-of-the-art CNN model for image denoising (i.e., DnCNN \cite{zhang2017beyond}).

\vspace{-0.25cm}
\subsection{Experimental settings}
\vspace{-0.15cm}
In the following experiments, we consider grayscale images. For training, we use the 432 training images of the BSD500 dataset \cite{arbelaez2011contour}. Instead, for testing we use a set of 12 widely used images. We train the network using a fixed noise standard deviation, considering three different noise levels $\sigma=15,25,50$. We subdivide the images into patches of size 32$\times$32 and train the network on 200k patches for 30 epochs. The non-local graph selects the 8 nearest neighboring pixels in terms of Euclidean distance between the hidden feature vectors, excluding the spatially adjacent pixels. The number of hidden features is 66 for all layers, except for the ones in the branches of the preprocessing block, for which is 22.

\vspace{-0.35cm}
\subsection{Quantitative and qualitative results}
\vspace{-0.1cm}
Table \ref{tab:results} shows the PSNR results on the 12 images of the test set. We can see that the proposed architecture outperforms the competing methods on most of the images and it provides the best average scores. In order to highlight the importance of the non-local filters, Table \ref{tab:zeronl} compares the proposed method with a network having the same architecture of the proposed model but employing only local neighbors instead of local and 8 non-local. The results show that the nonlocal model is indeed key to achieving the best possible performance. Notice that the 0-NN GraphCNN still falls short of the DnCNN-S model, which suggests that there is still room for improvement (e.g., adding more layers), leading to further gains also for the 8-NN GraphCNN. In addition to these quantitative results, we also show a qualitative comparison of the denoising methods in Figs. \ref{fig:04_denoised} and \ref{fig:06_denoised}. We can clearly see that the proposed method provides the best visual quality, recovering finer details and producing fewer artifacts. Lastly, we show in Fig. \ref{fig:recfield} the receptive field of a pixel for the first four graph-convolutional layers. It is interesting to note that the receptive field is adapted to the image characteristics, covering only a homogeneous region of the image. 

\vspace{-0.1cm}
\subsection{Comparison with other non-local approaches}
\vspace{-0.1cm}
An alternative approach, called NN3D, to include non-local information in CNNs for the denoising task has recently been proposed by Cruz et al. \cite{cruz2018nonlocality} by using a global post-processing stage based on a non-local filter after the output of a denoising CNN. This stage performs block matching and filtering over the whole image denoised by the CNN. The graph-convolutional method we presented in this paper exploits non-locality in a different way by constructing hierarchical non-local filters. Due to this, the method by Cruz et al. could also be used as a post-processor to our network to further increase the performance, as shown in Table \ref{tab:nn3d} (a single iteration of the NN3D method is used).

\begin{table}[]
\small
\centering
\vspace{-0.1cm}
\caption{Set12 average PSNR (dB) without non-local NN}
\vspace{-0.3cm}
\begin{tabular}{lccc}
& $\sigma = 15$ & $\sigma = 25$ & $\sigma = 50$ \\
\hline
DnCNN-S & 32.859 & 30.436 & 27.178 \\
\hline
GraphCNN (0-NN) & 32.757 & 30.348 & 27.048 \\
\hline
GraphCNN (8-NN) & 32.917 & 30.516 & 27.217 \\
\hline
\label{tab:zeronl}
\end{tabular}
\vspace{-0.3cm}
\end{table}

\begin{table}[]
\small
\setlength{\tabcolsep}{2pt}
\caption{Set12 average PSNR (dB) with NN3D}
\vspace{-0.2cm}
\begin{tabular}{lcccc}
$\sigma$ & DnCNN-S & DnCNN-S+NN3D & GraphCNN & GraphCNN+NN3D \\
\hline
25 & 30.436 & 30.448 & 30.516 & \textbf{30.517}\\
\hline
50 & 27.178 & 27.215 & 27.217 & \textbf{27.258}\\
\hline
\label{tab:nn3d}
\end{tabular}
\vspace{-0.6cm}
\end{table}

\vspace{-0.3cm}
\section{Conclusions and future work}
\label{sec:conclusions}
\vspace{-0.15cm}
In this paper we proposed a novel architecture of a convolutional neural network for image denoising that leverages graph-convolutional layers in order to create a hierarchy of non-local filters. Experimental results showed improved performance in terms of PSNR and visual quality. We can clearly attribute the gains to the non-local filters. Future work will focus on further improving the performance of the architecture. In particular, we will address the over-parameterization issues of graph convolution, which may unlock further gains.

\clearpage

\ninept
\bibliographystyle{IEEEbib}
\bibliography{biblio}

\end{document}